\documentclass[journal]{IEEEtran}

\usepackage{amsmath,amssymb}
\usepackage{cite}
\usepackage{float}
\usepackage{graphicx}
\usepackage{url}

\newtheorem{lemma}{Lemma}
\newtheorem{theorem}{Theorem}

\newcommand\ket[1]{\ensuremath{|#1\rangle}}
\newcommand\bra[1]{\ensuremath{\langle#1|}}

\newcommand\oprod[2]{\ensuremath{|#1\rangle\langle#2|}}
\newcommand\tr{\mathop{\rm tr}\nolimits}
\newcommand\var{\mathop{\rm Var}\nolimits}
\newcommand\re{\mathop{\rm Re}\nolimits}

\begin{document}

\title{Parameter estimation of quantum channels}

\author{Zhengfeng~Ji, Guoming~Wang, Runyao~Duan, Yuan~Feng and Mingsheng~Ying%
\thanks{This work was partly supported by the National Natural Science Foundation of China (Grant Nos.~60503001, 60421001 and 60621062) and the Hi-Tech Research and Development Program of China (863 project) (Grant No.~2006AA01Z102). Y. Feng was partially supported by the FANEDD under Grant No.~200755.}%
\thanks{Z. Ji is with the State Key Laboratory of Computer Science, Institute of Software, Chinese Academy of Sciences, P.O.Box 8718, Beijing 100080, China and formerly with the State Key Laboratory of Intelligent Technology and Systems, Department of Computer Science and Technology, Tsinghua University, Beijing 100084, China. E-mails: jzf@ios.ac.cn, jizhengfeng98@mails.tsinghua.edu.cn}%
\thanks{G. Wang, R. Duan, Y. Feng and M. Ying are with the State Key Laboratory of Intelligent Technology and Systems, Department of Computer Science and Technology, Tsinghua University, Beijing 100084, China. E-mails:
wgm00@mails.tsinghua.edu.cn (G. Wang),
dry@mail.tsinghua.edu.cn (R. Duan),
feng-y@mail.tsinghua.edu.cn (Y. Feng)
and yingmsh@mail.tsinghua.edu.cn (M. Ying)}}

\maketitle

\begin{abstract}
  The efficiency of parameter estimation of quantum channels is
  studied in this paper. We introduce the concept of programmable
  parameters to the theory of estimation. It is found that
  programmable parameters obey the standard quantum limit strictly;
  hence no speedup is possible in its estimation. We also construct a
  class of non-unitary quantum channels whose parameter can be
  estimated in a way that the standard quantum limit is broken. The
  study of estimation of general quantum channels also enables an
  investigation of the effect of noises on quantum estimation.
\end{abstract}


\section{Introduction}
\label{sec:intro}

\PARstart{P}{arameter} estimation, which is central to mathematical
statistics, is also an elementary problem in information theory. Its
main objective is to construct and evaluate various methods that can
estimate the values of parameters of either an information source or a
communication channel. Unlike in the usual scenarios of information
theory where the source and the channel are exactly known, we now have
a source or a channel that depends on some unknown parameters. Taking
the binary symmetric channel for example, we might know that the
channel is indeed binary symmetric but does not have any information
about the probability of it making a flip error. Thus, before we can
make use of it in communication, we should better determine the error
probability first. This is the most basic situation where parameter
estimation takes place and we will see later that it also arises in
other quite different applications.

Historically, the research of this topic dates back to the origin of
mathematical statistics, though the concept of ``a family of
distributions with parameters'' did not emerge until the 20's of the
last century~\cite{Chen02}. In the development, statisticians have
established different methods to make inferences about parameters:
maximum likelihood estimators, Bayes estimators, method of moments
estimators, etc. (see, for example,~\cite{Roh76} for detailed
discussions). At the same time, an important inequality---the
Cram\'{e}r-Rao inequality---was discovered which sets lower bounds on
the variance of any estimator in terms of Fisher
information~\cite{Rao45,Cra46,CT91}. Fisher further showed that
maximum likelihood estimators can achieve the lower bound
asymptotically~\cite{Fish22,Cra46} and made the Cram\'{e}r-Rao
inequality essential to estimation theory. These results from
statistics have already been applied to various problems in
information theory.

As quantum mechanics provides us with a more precise model of
describing reality, it is necessary to study estimation theory directly
based on quantum mechanics instead of the empirical models in
statistics. Helstrom~\cite{Hel68,Hel76} and Holevo~\cite{Hol82}
pioneered the study of estimation theory in the quantum setting. The
quantum version of Cram\'{e}r-Rao inequality was established
in~\cite{Hel68,Hel76,Hol82,BC94,BCM96}. It was shown by Braunstein,
Caves and Milburn that the lower bound, the reciprocal of quantum
Fisher information, is also achievable asymptotically~\cite{BCM96}.
This inequality has fundamental implications in physics. It is closely
related to skew information proposed by Wigner and
Yanase~\cite{Luo03,LZ04} and also implies the parameter-based version
of Heisenberg's uncertainty relation~\cite{BCM96}.

From the Cram\'{e}r-Rao inequality, or alternatively, from the central
limit theorem, we know that the standard deviation of an estimator
scales of order $1/\sqrt{N}$ where $N$ is the number of samples
observed from the parameterized source. Such a rate of convergence is
fundamental and universal. It also occurs in parameter estimation of
quantum information sources as pointed out, for example,
in~\cite{BCM96}. In the physics literature, the scaling of order
$\Omega(1/\sqrt{N})$ is sometimes called the standard quantum limit or
the shot noise limit. A fascinating aspect of the quantum case
is that such a limit can be beaten! Namely, if instead of estimating
parameters of a quantum information source, we are interested in
knowing to some precision parameters of a quantum channel, then it is
possible to have the scaling of $O(1/N)$ where $N$ stands for the
number of times the channel being used. The new scaling is the
so-called Heisenberg limit and accounts for a quadratic speedup in the
estimation compared to the standard quantum limit. This important
observation of fast estimation arises recently in a bunch of papers
which is motivated by applications in the most diverse fields: quantum
clock~\cite{BDM99}, clock synchronization~\cite{GLM01,BB05}, transfer
of reference frame~\cite{RG03,CDPS04,BBT04}, and so
on~\cite{Hay06,Kahn06,GLM06,Bal05,LBS+04,CPR00,NOB+07}. For a more complete
enumeration, see the recent survey papers~\cite{GLM04,BRS06}.

Parameter estimation of quantum channels is thus special: there are
parameters that can be estimated with a convergence rate never
achievable in the classical theory. We will call an estimator
\textit{superefficient\/} if it converges faster than the standard
quantum limit. Later on, we will also call a parameter superefficient
(inefficient) if it can (cannot) be estimated superefficiently. The
research of superefficient parameters of quantum channel is important
not only to the various applications arises in practice, but also to
the theory of quantum information and statistics. It fundamentally
characterizes the precision threshold that quantum mechanics permits
in a measurement.

However, most of the previous works focus only on fast parameter
estimation of unitary evolutions, the noiseless quantum channels. In
this paper, we will initiate the study of parameter estimation of a
general quantum channel, including the unitary transform. And the
emphasis of this paper is to characterize parameters that can or
cannot be estimated superefficiently.

Fast parameter estimation of unitary evolutions will be reviewed
briefly. We will analyze estimation protocols that can exceed the
standard quantum limit and will see the intrinsic relation of two
seemingly different protocols.

Next, we will provide a general criterion that rules out the
possibility of a large class of parameters being estimated
superefficiently. This is made possible by introducing the concept of
programmability to the estimation theory. If a family of channels
specified by some parameter is programmable, then any estimation
protocol of the parameter cannot exceed the standard quantum limit.
That is, programmable parameter of a quantum channel behaves much like
a classical one, unable to exploit the quantum advantage. The
programmability argument, though simple, has non-trivial implications
and extremely simplifies the analysis. For example, an interesting
corollary of it is that all parameters of classical discrete
memoryless channels are inefficient. Another important implication is
that the presence of depolarizing noise, no matter how small, will
``ruin'' the efficiency of estimation of all quantum channels.

On the other hand, we will also apply a general technique that can
help the construction of superefficient estimation protocols. This
technique is borrowed from Rudolph and Grover's method of establishing
a shared reference frame. Using this technique, we will show that
parameters in a non-unitary quantum channel may also be estimated
superefficiently.

This paper is organized as follows. Section~\ref{sec:background}
devotes to the introduction of some basic notations of estimation
theory and quantum information theory. We will discuss the
Cram\'{e}r-Rao inequality in both the classical and quantum setting in
this section. In Section~\ref{sec:unitary}, we review and analyze some
of the protocols that estimate parameter of unitary operations
superefficiently. A technique of parameter amplification is discussed in
detail which will be used later in Section~\ref{sec:projector_class}.
The concept of programmable channels is introduced in
Section~\ref{sec:programmability}. Some of the parameter estimation
problems and interesting corollaries are studied in this section based
on the ``no-go'' criterion we propose in terms of programmability.
Section~\ref{sec:projector_class} provides non-trivial examples of
fast parameter estimation of non-unitary channels.

\section{Notations and backgrounds of estimation theory and quantum
  information theory}
\label{sec:background}

In this section, we will discuss several topics that are important to
this work. First, we will review some of the basic facts of the
classical theory of parameter estimation. We will then move on to the
quantum case after a brief introduction to the concepts and notations
of quantum information theory.

\subsection{The classical theory of parameter estimation}
\label{sec:estimation}

In mathematical statistics, the parameter estimation problem is
formalized in terms of a family of distributions $f(x;\theta)$. Here
$\theta$ is the parameter to be estimated which belongs to a
\textit{parameter set\/} $\Theta$. We will only consider bounded
parameter set for simplicity. Suppose that a sample $\xi_1, \xi_2,
\ldots, \xi_N$ of size $N$ is drawn from the parameterized
distribution independently. An \textit{estimator\/} $\hat\theta$ for
$\theta$ for this sample is a function of the $N$ observed values
$\hat\theta(\xi_1, \xi_2, \ldots, \xi_N)$ valued in $\Theta$.

The estimator is said to be \textit{unbiased\/} if its expectation
$E(\hat\theta)$ equals to the unknown parameter $\theta$. Another
qualitative evaluation of an estimator is its consistency: we say an
estimator is \textit{consistent\/} if it converges to the unknown
parameter in probability as the sample size tends to infinite. We
would only consider consistent estimators in this paper. To evaluate
an estimator $\hat\theta$ quantitatively, the \textit{mean squared
  error\/} (MSE)
\begin{equation}
  \label{eq:mse}
  E (\hat\theta - \theta)^2
\end{equation}
is usually employed. The smaller the MSE, the better
precision the estimator promises.

In the language of information theory, the distributions $f(x;\theta)$
can be thought of as the statistics of a memoryless source with a
hidden parameter $\theta$. For example, it can be a discrete
memoryless source $\mathcal{B}_\theta$ with source statistics $p(0) =
1-\theta$, $p(1) = \theta$ where $\theta \in [0, 1]$ is the parameter.
A good estimator for $\theta$ one can easily imagine is the sample
mean $\bar\theta = \sum_i \xi_i/N$. The estimator is obviously
unbiased and has a variance of $\theta(1-\theta)/N$. For an unbiased
estimator, the mean squared error is equal to its variance. Thus, we
would like to find an unbiased estimator with as small variance as
possible. However, the Cram\'{e}r-Rao inequality sets lower bounds on
the variance. For example, it tells us that $\bar\theta$ has the least
variance possible among all unbiased estimators of $\theta$.

\begin{theorem}[The Cram\'{e}r-Rao inequality]
For all estimator $\hat\theta(\xi)$,
\begin{equation}
 \label{eq:biased_crb}
 \var \hat\theta \ge {\bigl(dE(\hat\theta)/d\theta\bigr)^2 \over
   J(\theta)}.
\end{equation}
Here $J(\theta)$ is the Fisher information defined as
\begin{equation}
J(\theta) = E\Bigl[{\partial\over\partial\theta} \ln f(X;\theta) \Bigr]^2,
\end{equation}
where $X \sim f(x;\theta)$.
\end{theorem}

When $\hat\theta$ is unbiased, $dE(\hat\theta)/d\theta = 1$, so we can
rewrite the inequality as
\begin{equation}
 \label{eq:unbiased_crb}
 E \bigl(\hat\theta - \theta\bigr)^2 \ge {1 \over J(\theta)}.
\end{equation}

The proof of the above theorem can be found in, for
example,~\cite{CT91} or~\cite{BC94}. It is also easy to show that
Fisher information $J(\theta)$ is additive. Concretely, let
$J_1(\theta)$, $J_2(\theta)$ be the Fisher information of distributions
$f(x;\theta)$ and $g(y;\theta)$ respectively. The Fisher information
$J_{12}(\theta)$ of the joint distribution $f(x;\theta)g(y;\theta)$ is
equal to $J_1(\theta) + J_2(\theta)$. Applying this observation, we
can get the Cram\'{e}r-Rao inequality for estimators of sample size
$N$:
\begin{equation}
 \label{eq:unbiased_crb_N}
 E \bigl(\hat\theta - \theta\bigr)^2 \ge {1 \over N J(\theta)}.
\end{equation}
It can be easily verified that for source $\mathcal{B}_\theta$, the
Fisher information $J(\theta)$ is $[\theta(1-\theta)]^{-1}$. Thus
$\bar\theta$ is optimal as mentioned. In some cases, it might be
possible that there does not exist any estimator that can saturate the
lower bound in the Cram\'{e}r-Rao inequality. However, Fisher showed
that, except for some extreme cases, the maximum likelihood estimator
can always achieve the lower bound in the limit of large sample size
$N$~\cite{Fish22}.

A corollary of Eq.~\eqref{eq:unbiased_crb_N} which is important to
this paper is that no unbiased estimator can have its variance
converging to zero at a rate faster than the order of $1/N$, where $N$
is the sample size. In terms of the standard deviation, this is a
convergence rate of $\Omega(1/\sqrt{N})$. In the following, we will
call an estimator, or an estimation protocol, is of order $1/\sqrt{N}$
($1/N$, etc.) if its standard deviation converges with order
$1/\sqrt{N}$ ($1/N$, resp.) for all possible $\theta$.

In the previous analysis, we have derived the rate of convergence from
Eq.~\eqref{eq:unbiased_crb_N} which only applies to unbiased
estimators. We now claim that biased estimators are also of order
$\Omega(1/\sqrt{N})$ in terms of the root mean squared error (RMSE)
instead of the standard deviation. As we have assumed the parameter
set to be bounded, $E(\hat\theta)$ converges point-wise to $\theta$ as
$N \rightarrow \infty$ for any consistent estimator $\hat\theta$. It
follows from the mean value theorem that there exists a specific
$\theta_0$ such that
\begin{equation*}
  {dE(\hat\theta) \over d\theta}\bigg|_{\displaystyle\theta_0}
\end{equation*}
is close to $1$ for large $N$. Combining the fact that
\begin{equation*}
  E(\hat\theta-\theta)^2 \ge \var\hat\theta
\end{equation*}
and Eq.~\eqref{eq:biased_crb}, we complete the justification of the
claim. Thus, we have shown that any estimator of parameters of a
classical information source is of order $\Omega(1/\sqrt{N})$. We note
that the locally normalized deviation measure
\begin{equation}
  E\Bigl(\frac{\hat\theta}{dE(\hat\theta)/d\theta} - \theta\Bigr)^2
\end{equation}
was employed to deal with the case of biased estimators
in~\cite{BC94}. We insist on using MSE in this paper as it is much
easier to calculate and provides us with a uniform criterion in
evaluating different estimators.

Before we introduce the quantum Cram\'{e}r-Rao inequality, we will
first review quantum mechanics form an information-theoretical point
of view. For a more detailed presentation of the theory of quantum
information, the readers are referred to~\cite{NC00}.

\subsection{Quantum information sources and quantum channels}
\label{sec:quantum}

In quantum information theory, quantum state plays the role of the
information carrier. Any \textit{quantum state\/} can be described by
a positive semidefinite operator $\rho$ with unit trace. When
diagonal, it degenerates to a discrete probability distribution and is
thus also a natural description of a quantum information source.

The evolution of a closed quantum system is characterized by a unitary
operation $U$ which maps $\rho$ to $U\rho U^{\dagger}$. As a special
type of quantum channel, unitary evolution is invertible and
noiseless. A general \textit{quantum channel\/} is mathematically a
superoperator $\mathcal{E}$ which is completely positive and
trace-preserving. That is, for any positive semidefinite operator
$\rho$, $\mathcal{I}\otimes\mathcal{E} (\rho)$ is positive
semidefinite and $\tr \bigr(\mathcal{E(\rho)}\bigl)= \tr(\rho)$ where
$\mathcal{I}$ is the identity superoperator. The effect of any quantum
channel $\mathcal{E}$ can be viewed as the dynamics of one part of a
larger closed system. Namely, there always exists a unitary operation
$U$ such that for all $\rho$,
\begin{equation}
  \label{eq:traced_u}
  \mathcal{E} (\rho) = \tr_{env} \left[
    U(\rho\otimes\oprod{0}{0}_{env})U^{\dagger}\right].
\end{equation}
Another description of quantum channels which is easy to use is the
Kraus' \textit{operator-sum representation\/}. In this representation, any
channel $\mathcal{E}$ is specified by a set of operators $E_i$
satisfying $\sum_i E_i^{\dagger}E_i = I$, and
\begin{equation}
 \mathcal{E}(\rho) = \sum_i E_i \rho E_i^{\dagger}.
\end{equation}
Different sets of operators, $\{E_i\}_{i=1}^n$ and $\{F_j\}_{j=1}^m$,
may correspond to the same quantum channel. When $m=n$, this occurs if
and only if there exists $u_{ij}$ such that $E_i = \sum_j u_{ij} F_j$
and $(u_{ij})$ is unitary. It is thus called the unitary freedom in
the operator-sum representation~\cite{NC00}. Note that in the case of
$m\ne n$, we can append zero operators to the set having the smaller
number of operators.

One of the simplest quantum channels of interest is the qubit
depolarizing channel
\begin{equation}
  \mathcal{E}(\rho) = p \frac{I}{2} + (1-p) \rho.
\end{equation}
It is naturally the quantum counterpart of the binary symmetric
channel. One of its operator-sum representations is specified by
\begin{equation}
  \label{eq:depolarizing_operators}
  \left\{\sqrt{1-3p/4} I, \sqrt{p}X/2, \sqrt{p}Y/2, \sqrt{p}Z/2 \right\}
\end{equation}
where $X,Y,Z$ are the Pauli matrices. The Pauli matrices may also be
denoted by $\sigma_i$'s sometimes:
\begin{equation}
  \label{eq:pauli}
  \begin{split}
  I = \sigma_0 = \begin{bmatrix}1&\phantom{-}0\\0&\phantom{-}1\end{bmatrix},\quad
  X = \sigma_1 = \begin{bmatrix}0&\phantom{-}1\\1&\phantom{-}0\end{bmatrix},\\
  Y = \sigma_2 = \begin{bmatrix}0&-i\\i&\phantom{-}0\end{bmatrix},\quad
  Z = \sigma_3 = \begin{bmatrix}1&\phantom{-}0\\0&-1\end{bmatrix}.
  \end{split}
\end{equation}

A special type of non-unitary quantum operation which is important to
the interpretation of quantum theory is quantum measurements. Quantum
measurement is the bridge that links the quantum and classical worlds
and is the only way for us to obtain classical information from a
quantum system. One of the formulations of \textit{quantum
  measurements\/} is described by the resolution of identity $I$ into
projectors $P_i$'s, $I = \sum_i P_i$. The probability of observing $k$
is $\tr(\rho P_k)$ and the post-measurement state becomes $P_k\rho
P_k/\tr(\rho P_k)$. If we do not care much about the post-measurement
state, we can employ another description called
\textit{positive-operator valued measure (POVM)}. Mathematically, it
is a resolution of identity $I$ into positive semidefinite operators
$M_i$, $I = \sum_{i=1}^m M_i$. The probability of observing result $k$
is $\tr(\rho M_k)$. For example, the measurement along the basis
$\ket{+} = (\ket{0}+\ket{1})/\sqrt{2}$ and $\ket{-} = (\ket{0} -
\ket{1})/\sqrt{2}$ can be modeled by $P_+$ and $P_-$,
\begin{equation}
  \begin{split}
  P_+ = \oprod{+}{+} & =
  \frac{1}{2} \begin{bmatrix}1&1\\1&1\end{bmatrix},\\
  P_- = \oprod{-}{-} & =
  \frac{1}{2} \begin{bmatrix}\phantom{-}1&-1\\-1&\phantom{-}1\end{bmatrix}.
  \end{split}
\end{equation}

Measurements are quantum channels. Thus, they can also be described by
the operator-sum representation. The above simple example can be
specified by the following set of operators
\begin{equation}
  \left\{ \oprod{0}{+}, \oprod{1}{-} \right\}.
\end{equation}
Notice that we have chosen the measurement result, instead of the
post-measurement state, to be the outcome of the channel.

\subsection{The quantum Cram\'{e}r-Rao inequality}
\label{sec:quantum_cri}

We are now ready to introduce the quantum Cram\'{e}r-Rao inequality
which first appeared in~\cite{Hel68}. We will sketch the proof for it
because of its importance to one of our results. The proof is similar
to the one given in~\cite{BC94}.

Consider a quantum information source $\rho(\theta)$ which depends on
parameter $\theta$. It is beneficial to divide an estimation protocol
into two different steps~\cite{BC94}. In the first step, perform a
properly designed measurement $M$, and in the second, make an
estimation based on the data obtained in the previous step. On can see
that the second step is essentially the same as a classical estimation
protocol and the classical Cram\'{e}r-Rao inequality applies. That is,
given the POVM $M = \{M_i\}_{i=1}^m$ chosen in the first step, we get
a lower bound that depends on $M$
\begin{equation}\label{eq:crb_m}
  E(\hat\theta-\theta)^2 \ge \frac{1}{J_M(\theta)},
\end{equation}
where
\begin{equation}
  J_M(\theta) = \sum_{i=1}^m \frac{\bigl[\tr(M_i\rho')\bigr]^2}{\tr(M_i\rho)}.
\end{equation}
We have considered only unbiased estimators here and the biased case
can be analyzed similarly as in the classical case.

Write the spectrum decomposition $\rho = \sum_i p_i\oprod{i}{i}$ and
define a superoperator $\mathcal{L}_\rho$ as
\begin{equation}
  \label{eq:L_rho}
  \mathcal{L}_\rho (O) = \sum_{\{j,k \mid p_j+p_k\ne 0\}}
  \frac{2}{p_j+p_k} O_{jk}\oprod{j}{k}.
\end{equation}
An important property of $\mathcal{L}_\rho$ is that for non-singular
$\rho$, and Hermitian matrices $A$ and $B$,
\begin{equation}
  \label{eq:L_rho_prop}
  \tr (AB) = \re \bigl[ \tr(\rho A \mathcal{L}_\rho(B)) \bigr].
\end{equation}

It follows by substitution that
\begin{equation}
  J_M(\theta) = \sum_{i=1}^m \frac{\Bigl(\re\bigl[\tr\bigl(\rho
    M_i\mathcal{L}_\rho(\rho')\bigr)\bigr]\Bigr)^2}{\tr(M_i\rho)}.
\end{equation}
The validity of this substitution for singular $\rho$ is justified
in~\cite{BC94}. Hence,
\begin{equation}\label{eq:bound_j_m}
\begin{split}
  J_M(\theta) & \le \sum_{i=1}^m \frac{\bigl|\tr\bigl(\rho
    M_i\mathcal{L}_\rho(\rho')\bigr)\bigr|^2}{\tr(M_i\rho)}\\
              & = \sum_{i=1}^m \Biggl| \tr \Bigl( {\rho^{1/2}M_i^{1/2} \over
                \sqrt{\tr(M_i\rho)}} M_i^{1/2} \mathcal{L}_\rho(\rho')
              \rho^{1/2}\Bigr) \Biggr| ^2\\
              & \le \sum_{i=1}^m \tr \bigl( M_i \mathcal{L}_\rho(\rho')
              \rho \mathcal{L}_\rho(\rho') \bigr)\\
              & = \tr \bigl( \mathcal{L}_\rho(\rho') \rho
              \mathcal{L}_\rho(\rho') \bigr)\\
              & = \tr \bigl( \rho' \mathcal{L}_\rho(\rho') \bigr),
\end{split}
\end{equation}
where the second inequality follows from the Cauchy-Schwarz inequality.

The term in the final step of Eq.~\eqref{eq:bound_j_m} is the quantum
Fisher information
\begin{equation}
  \label{eq:quantum_Fisher}
  J(\theta) = \tr \bigl( \rho' \mathcal{L}_\rho(\rho') \bigr),
\end{equation}
which depends only on the parameterized state $\rho$ and we may also
denote it by $J_{\rho}(\theta)$ for clarity.

The following theorem follows from Eqs.~\eqref{eq:crb_m}
and~\eqref{eq:bound_j_m}.

\begin{theorem}[The quantum Cram\'{e}r-Rao inequality]
For any unbiased estimator $\hat\theta$ for $\theta$ of $\rho(\theta)$,
\begin{equation}
  \label{eq:quantum_crb}
  E(\hat\theta-\theta)^2 \ge {1 \over J(\theta)}.
\end{equation}
\end{theorem}

Next, we show that the quantum Fisher information is also additive.
That is,
\begin{equation}\label{eq:quantum_fisher_additive}
  J_{\rho}(\theta) = J_{\sigma}(\theta) + J_{\tau}(\theta),
\end{equation}
if $\rho = \sigma \otimes \tau$.
The proof is simple. As $\rho' = \sigma'\otimes\tau + \sigma\otimes\tau'$,
\begin{equation}
  \begin{split}
    \tr\bigl(\rho'\mathcal{L}_\rho(\rho')\bigr) & = \tr \bigl(
    \sigma'\mathcal{L}_{\sigma}(\sigma') \otimes \tau +
    \sigma'\otimes\tau\mathcal{L}_{\tau}(\tau') \\
    & + \sigma \otimes \tau'\mathcal{L}_{\tau}(\tau') +
    \sigma\mathcal{L}_{\sigma}(\sigma')\otimes\tau'\bigr),
  \end{split}
\end{equation}
and Eq.\eqref{eq:quantum_fisher_additive} follows by noticing that
$\tr(\sigma') = \tr(\tau') = 0$.

Therefore, if $N$ replicas of $\rho$, $\rho^{\otimes N}$, is used in
the estimation, we have the corresponding Cram\'{e}r-Rao inequality
\begin{equation}
  E(\hat\theta-\theta)^2 \ge {1 \over N J(\theta)}.
\end{equation}
This means that any unbiased estimator is also of order
$\Omega(1/\sqrt{N})$ and so is the biased case by a similar argument
used before. We would also like to point out that the above analysis
applies to any joint measurement on the $N$ copies as mentioned
in~\cite{BCM96}. Thus, the standard quantum limit is essential for all
parameters of quantum information sources. We will refer to this
result later in Section~\ref{sec:programmability}.

\section{A review of parameter estimation for unitary operations}
\label{sec:unitary}

Unitary evolution is one of the most fundamental operations in quantum
information, and has therefore received the most attentions. It is
also the first type of operations studied in parameter estimation of
quantum channels. The most amazing observation is that, unlike
parameters of both classical and quantum information sources,
parameters of unitary operations can be estimated much faster! Two
different approaches of superefficient estimation are studied in the
following. One of them is of order $1/N$; the other is of order $\log
N/N$.

\subsection{Strategies that can beat the standard quantum limit}
\label{sec:beat}

The parameter $\theta$ now determines a unitary $U(\theta)$. To
estimate the value of $\theta$, we will apply the unitary $N$ times to
some states properly prepared. Then, the problem becomes parameter
estimation of states which we are more familiar with. However, it is
much more flexible to employ an operation in an estimation protocol:
it can be carried out in parallel, sequentially or even by mixing both
of the two. In the following, we will discuss some of the superefficient
strategies that are of common use.

\begin{figure}[!ht]
  \centering
  \includegraphics{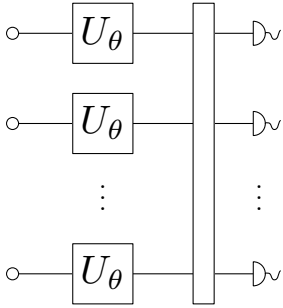}
  \caption{Layout of the parallel strategy}
  \label{fig:parallel}
\end{figure}

The first simple strategy we discuss carries out the unitaries in
parallel. See Fig.~\ref{fig:parallel} for a demonstration of the
layout. It will be referred to as the parallel strategy. Now, if the
input are chosen to be product states of $\rho_1$, $\rho_2$, \dots,
$\rho_N$, we will show that no estimation protocol can beat the
standard quantum limit no matter what kind of joint measurement and
the post-measurement estimator one chooses. This is seen by the
additivity of quantum Fisher information and the quantum
Cram\'{e}r-Rao inequality.

\floatstyle{plain}
\restylefloat{figure}

Before the measurement, the state can be written as
\begin{equation}
  \mathop{\bigotimes}\limits_{i=1}^N \left[
    U(\theta)\rho_iU^{\dagger}(\theta) \right],
\end{equation}
whose Fisher information of $\theta$ is
\begin{equation}
  \sum_{i=1}^N J_{U\rho_iU^{\dagger}}(\theta) \le N \max_{\rho} \left\{J_{U\rho
    U^{\dagger}}(\theta)\right\}.
\end{equation}
The convergence rate of $\Omega(1/\sqrt{N})$ follows immediately from
the quantum Cram\'{e}r-Rao inequality. This result is also noted
in~\cite{GLM06} as the so-called CC and CQ strategies considered
there.

If the input of the $N$ parallel unitary operations is chosen to be
some entangled state, the argument above does not work anymore. In
fact, we can find estimations of order $O(1/N)$ with the help of
quantum entanglement.

To make the analysis simpler, we will focus on the estimation of
$\theta\in\Theta=[0,1)$ of a single-qubit unitary
\begin{equation}
  \label{eq:single_phase}
  U = \begin{bmatrix}1 & 0\\0 & e^{2\pi i\theta}\end{bmatrix}.
\end{equation}
Yet, we claim that this simple case is essentially as general as the
estimation of the angular parameter of $U = e^{-i\theta H}$ where $H$
is a known Hermitian operator independent of $\theta$. In the basis of
eigenvectors of $H$, $U$ has a diagonal matrix representation and
operates as a single-qubit unitary defined in
Eq.~\eqref{eq:single_phase} when restricted to a two dimensional
subspace. We therefore do not lose much by confining our attention to
$U$ defined in Eq.~\eqref{eq:single_phase}. The estimation of angular
parameter, though simple, has wide applications in
physics~\cite{Hol82,GLM06,BDM99}.

\begin{figure}[!ht]
  \centering
  \includegraphics{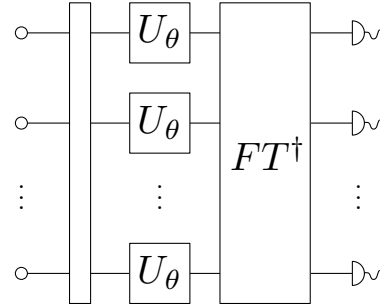}
  \caption{Layout of the parallel strategy with entangled inputs and
    Fourier basis measurement}
  \label{fig:parallel_fft}
\end{figure}

In Holevo's book~\cite{Hol82}, the optimal estimation of parallel
strategies was found for angular parameter of $U = e^{-i\theta H}$
based on a theory of covariant measurements. The result was employed
recently by Bu\v{z}ek, Derka and Massar in designing optimal quantum
clocks~\cite{BDM99} which can achieve the Heisenberg limit. This
speedup was found by optimizing the input state that is fixed in
Holevo's result. We will present an analysis which is similar to the
one given by Hayashi~\cite{Hay06} but will appeal to the Fourier basis
measurement instead of the covariant measurement. The Fourier basis
measurement is in fact one of the discrete versions of the covariant
measurement~\cite{DBE98}. The procedure is illustrated in
Fig.~\ref{fig:parallel_fft}.

\floatstyle{plain}
\restylefloat{figure}

Consider $U(\theta)$ given in Eq.~\eqref{eq:single_phase} and define $N+1$
special states in the space on which $U^{\otimes N}$ acts:
\begin{equation}
  \ket{\hat k} = \frac{1}{\sqrt{N \choose l}}\sum_{l:w(l)=k} \ket{l},
\end{equation}
for $k=0,1,\ldots,N$ where $w(l)$ is the Hamming weight of $l$.
By virtue of the parallel structure, we have
\begin{equation}
  U^{\otimes N} \ket{\hat k} = e^{2k\pi i \theta} \ket{\hat k}.
\end{equation}
This means that $U^{\otimes N}$ will rotate the input state
\begin{equation}\label{eq:pre_u_state}
  \sum_{k=0}^N a_k \ket{\hat k},\quad a_k \text{ is real}
\end{equation}
to
\begin{equation}\label{eq:post_u_state}
  \sum_{k=0}^N a_k e^{2k\pi i\theta}\ket{\hat k},
\end{equation}
where the $a_k$'s will be given later. The inverse Fourier transform
\begin{equation}
  \ket{\hat k} \mapsto \frac{1}{\sqrt{N+1}}\sum_{l=0}^N e^{-2\pi i
    kl/(N+1)}\ket{\hat l}
\end{equation}
brings the state further to
\begin{equation}
  \frac{1}{\sqrt{N+1}} \sum_{k,l=0}^N a_ke^{2k\pi
    i(\theta-l/(N+1))}\ket{\hat l}.
\end{equation}
Finally, perform the computational basis measurement on each of the
qubit, and estimate $\theta$ with the number of $1$'s in the
measurement outcome divided by $N+1$. This completes the description
of the parallel strategy.

What remains to be clarified is the efficiency of this protocol. We
will choose the expectation, denoted by $W$, of
$1-\cos(2\pi(\hat\theta-\theta))$ instead of the MSE. One can see that
these two evaluations are equivalent as
$1-\cos(2\pi(\hat\theta-\theta))$ is asymptotically $2\pi^2
(\hat\theta-\theta)^2$ when $(\hat\theta-\theta)$ is small. The choice
of this type of evaluation function, which is first used in Holevo's
book~\cite{Hol82}, helps to simplify the calculation of a close form
formula of the averaged deviation.

As the probability of observing $l$ $1$'s in the output is
\begin{equation}\label{eq:fft_prob}
  \begin{split}
  \Pr(l) & = \left| \sum_{k=0}^N a_ke^{2k\pi i(\theta-l/(N+1))}
  \right|^2 \Big/(N+1)\\
        & = \frac{1}{N+1}\sum_{m,n=0}^Na_ma_ne^{2\pi
          i(\theta-l/(N+1))(m-n)},
  \end{split}
\end{equation}
the expectation
\begin{equation}
  \begin{split}
    W & = E\bigl(1-\cos(2\pi(\hat\theta-\theta))\bigr)\\
      & = \sum_{l=0}^N \Pr(l) \Bigl[ 1 - \cos \bigl(
      2\pi(\theta-\frac{l}{N+1}) \bigr) \Bigr]\\
      & = 1 - \sum_{l=0}^N \Pr(l) \cos \bigl(
      2\pi(\theta-\frac{l}{N+1}) \bigr).
  \end{split}
\end{equation}
Employing Eq.~\eqref{eq:fft_prob} and the fact the LHS of the above
equation is real, we can continue the calculation as
\begin{equation}
  \begin{split}
    W & = 1 - \frac{1}{N+1} \re \sum_{l,m,n=0}^N a_ma_ne^{2\pi
        i(\theta-l/(N+1))(m-n+1)}\\
      & = 1 - \sum_{k=1}^N a_{k-1}a_k  - a_0a_N\cos(2\pi(N+1)\theta).
  \end{split}
\end{equation}
If we choose $a_0=0$, $W$ will be independent of $\theta$,
\begin{equation}
  W = 1 - \sum_{k=2}^N a_{k-1}a_k.
\end{equation}
Now, we need to minimize $W$ subject to the normalization condition
\begin{equation}
  \sum_{k=1}^N a_k^2 = 1.
\end{equation}
One can see that the minimum value of $W$ is equal to the minimum
eigenvalue the $N$ by $N$ matrix $A$ whose diagonal elements are all
$1$ and secondary diagonal elements are all $-1/2$:
\begin{equation}
  \renewcommand\arraystretch{1.5}
  A = \begin{bmatrix}
    1 & -\frac{1}{2} & & &\\
    -\frac{1}{2} & 1 & -\frac{1}{2} & &\\
    & -\frac{1}{2} & 1 & \ddots &\\
    & & \ddots & \ddots & -\frac{1}{2}\\
    & & & -\frac{1}{2} & 1
  \end{bmatrix}.
\end{equation}
The minimum eigenvalue
of $A$ is $2 \sin^2\frac{\pi}{2N+2}$ and the corresponding eigenvector
gives the values of $a_k$'s for $k\ge 1$:
\begin{equation}
  a_k = \sqrt{\frac{2}{N+1}} \sin \frac{k\pi}{N+1}.
\end{equation}

For large $N$, $W = 2 \sin^2\frac{\pi}{2N+2}$ is obviously of order
$1/N^2$. Thus the estimation protocol given above is of order $O(1/N)$
in terms of RMSE.

\begin{figure}[!ht]
  \centering
  \includegraphics{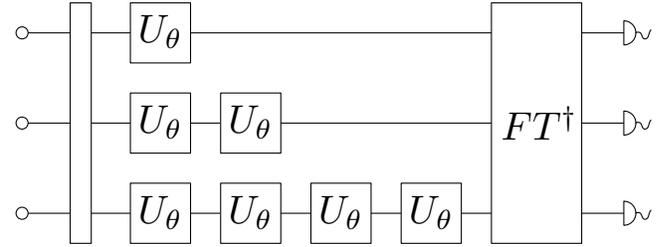}
  \caption{Layout of a mixed strategy based on Fourier transform}
  \label{fig:fft}
\end{figure}

Our next strategy mixes both parallel and sequential parts but still
has a simple structure as depicted in Fig.~\ref{fig:fft}. Namely, it
prepares $n$ qubits in parallel and applies the unitary $U$ on the $j$th
qubit $2^{j-1}$ times and thus $N = 2^n-1$ times in total. It is easy
to see that before the inverse Fourier transform, the state of the $n$
qubits is
\begin{equation}
  \sum_{k=0}^N a_k e^{2k\pi i\theta}\ket{k},
\end{equation}
given that the initial state is
\begin{equation}
  \sum_{k=0}^N a_k \ket{k}.
\end{equation}
The above two equations are have the same form of
Eqs.~\eqref{eq:pre_u_state} and~\eqref{eq:post_u_state}. Thus the
analysis of the mixed strategy presented here will be essentially the
same as the parallel strategy though they look quite different. It is
worth noting that, recently, a similar strategy by optimizing the
input state is discovered independently in~\cite{DDE+06} which
dramatically improves the average efficiency of the phase estimation protocol
proposed in~\cite{CEMM98}.

\floatstyle{plain}
\restylefloat{figure}

We have now seen how an estimation protocol beats the standard quantum
limit. A more ambitious question is whether it is possible to find
even better protocols which converge faster than the order of $1/N$.
Unfortunately, it has been proven impossible, for example,
in~\cite{GLM06} by employing an uncertainty relation implied by the
quantum Cram\'{e}r-Rao inequality. Though the proof there considers
only unitary operations, the result applies to general quantum
channels because of Eq.~\eqref{eq:traced_u}. The scaling of $1/N^k$ is
reported recently in a quite different problem setting~\cite{BFCG07}
and makes no contradictions.

Unlike the previous strategies we have discussed, another important
class of the strategies has much looser structures. The spirit of it
is trying to accumulate the parameter before we observe. In respect
that this class of strategies is closely related to one of the main
result of this paper, we organize the treatment of it in a separate
part.

\subsection{The technique of amplifying parameters}
\label{sec:amplify}

It has been widely noticed that angular parameters of a unitary can be
easily accumulated either with or without
entanglement~\cite{RG03,BB05,GLM04,GLM06}. Again, let $U$ be
\begin{equation}
  \label{eq:single_phase_2}
  \begin{bmatrix}1 & 0\\0 & e^{2\pi i\theta}\end{bmatrix}
\end{equation}
but consider only $\theta \in \Theta=[0,1/2]$ for simplicity. By
applying it in parallel to the $n$-qubit GHZ state~\cite{GHZ89}
\begin{equation}
  \label{eq:ghz}
  \frac{\ket{00\cdots 0} + \ket{11\cdots 1}}{\sqrt{2}},
\end{equation}
one gets
\begin{equation}
  \frac{\ket{00\cdots 0} + e^{2\pi i n\theta}\ket{11\cdots 1}}{\sqrt{2}},
\end{equation}
which is a state that depends on $n\theta$. The parameter is thus
amplified.

Angular parameters can also be amplified without entanglement. To see
this, we employ a sequential strategy that applies the same $U$ $n$
times on state $(\ket{0}+\ket{1})/\sqrt{2}$. The amplification is also
done since the state is finally rotated to
$(\ket{0}+ e^{2\pi i n\theta}\ket{1})/\sqrt{2}$.

Any estimation of $N\theta$ provides also an estimator for $\theta$
simply by dividing the estimated value by $N$. Moreover, if the
estimation of $N\theta$ has RMSE bounded by constant, the estimation
of $\theta$ seems to be of order $1/N$. However, it is noticed in the
literature that this argument is not rigorous~\cite{BB05} and only
applies for $\theta$ small enough (compared to $1/2N$). The reason is
that $e^{2\pi i N\theta}$ is periodical and we cannot always decide
the value of $\theta$ from $e^{2\pi i N\theta}$.

Fortunately however, there is an ingenious way that can deal with this
problem. Rudolph and Grover proposed an iterative procedure which can
determine the first $k$ bits of the parameter~\cite{RG03}. They used
this method to establish a shared reference frame between two remote
parties. Later, Burgh and Bartlett used it in their clock
synchronization protocols~\cite{BB05}. This technique, which can be
proven to be of order $O(\log N/N)$, is in fact a general method for
parameter estimation provided that the parameter can be amplified.

Yet, there is a loophole in Rudolph and Grover's bitwise protocol
which makes the protocol problematic sometimes. In their paper, $T$
and $T'$ are the parameter to be estimated and a possible estimation
respectively. They assume that $|T-T'|\le 1/4$ implies that $T'$
agrees with $T$ to at least the first bit. This is generally not true
no matter how close $T'$ and $T$ are. A careful verification tells
that if the true parameter $T$ is $1/2$, their protocol is not even a
consistent estimation, that is, $T'$ does not converge to $T$ in
probability. We will modify the protocol to close the loophole.

The modified protocol still contains $k$ steps. In the first step, we
prepare state $(\ket{0}+\ket{1})/\sqrt{2}$, apply $U$ once, and
measure along the Hadamard basis $\ket{+},\ket{-}$. The probability of
observing $+$ is $P_{+} = \cos^2(\pi\theta)$. Repeat the procedure $n$
times and calculate the sample mean $\bar P$ as an estimation for the
value of $P_{+}$. Let the parameter corresponding to $\bar P$ be
$\bar\theta$, that is, $\bar P = \cos^2(\pi\bar\theta)$. Obviously,
there exists some constant $\delta$ such that $|P_{+} - \bar P|\le
\delta$ implies $|\theta - \bar\theta| \le 1/12$. Choose $n$ large
enough to insure
\begin{equation}
  \label{eq:delta_prob}
  \Pr [ |P_{+} - \bar P| \le \delta ] \ge 1 - \epsilon/k.
\end{equation}
Thus, with at least the same probability, $|\theta - \bar\theta| \le
1/12$. Consider the following three cases depending on the value of
$\bar\theta$ calculated.
\begin{enumerate}
\item If $\bar\theta\in[0,5/12)$, the probability of
  $\theta\in[0,1/2]$ is at least $1 - \epsilon/k$. Define $r_1=2$ and
  $\nu_1=0$ in this case.
\item If $\bar\theta\in[5/12,7/12]$, the probability of
  $\theta\in[1/3,2/3]$ is at least $1 - \epsilon/k$. Define $r_1=3$,
  $\nu_1=1$.
\item Otherwise, $\bar\theta\in[7/12,1]$, the probability of
  $\theta\in[1/2,1]$ is at least $1 - \epsilon/k$. Define $r_1=2$,
  $\nu_1=1$.
\end{enumerate}
This finishes the first step.

In the first step we have insured that the true parameter belongs to
an interval of length $1/r_1$ with high probability. We would continue
this idea in the following steps. In the second step, we still prepare
$(\ket{0}+\ket{1})/\sqrt{2}$, but apply $U$ $r_1$ times instead, where
$r_1$ is determined in the previous step whose value is either $2$ or
$3$. The following is similar to the first step if we regard the
decimal part of $r_1\theta$ as $\theta$. The second step determines
the value of $r_2\in\{2,3\}$ and $\nu_2$ in a similar way. After the
second step, the ``possible'' interval of the true parameter is of
length $1/r_1r_2$. In the third step, the unitary $U$ is carried out
sequentially $r_1r_2$ times each trial. Similarly, $U$ is applied
$\prod_{i=1}^{k-1}r_i$ times each trial in the $k$th step. After all
the $k$ steps, we can make sure that the parameter $\theta$ is in an
interval of length $1/\prod_{i=1}^k r_i$ with probability at least
$(1-\epsilon/k)^k$. We conclude the whole procedure by accepting
\begin{equation}
  \hat\theta = \sum_{i=1}^k \Bigl( \nu_i \prod_{j=1}^i r_j^{-1} \Bigr)
\end{equation}
as the estimated value, represented in a mixed radix system.

To ensure Eq.~\eqref{eq:delta_prob}, we can choose
\begin{equation}
  n \ge \frac{1}{2 \delta^2} \ln 2k/\epsilon,
\end{equation}
which follows from the Chernoff inequality
\begin{equation}
  \label{eq:chernoff}
  \Pr[|P-\bar P| \ge \delta] \le 2 e^{-2 n \delta^2}.
\end{equation}

Thus the total number of times that $U$ is applied in the $k$ steps is
\begin{equation}
  \label{eq:total_number}
  N = n (1 + r_1 + r_1r_2 + \cdots + r_1r_2\cdots r_{k-1}).
\end{equation}

We will prove that the protocol given above is indeed an superefficient
estimation of order $O(\log N/N)$. After all the $k$ steps, the
probability
\begin{equation}
  \Pr\Bigl[|\hat\theta-\theta| \le 1\big/\prod_{i=1}^k r_i \Bigr] \ge
  (1-\epsilon/k)^k \ge 1 - \epsilon.
\end{equation}
Thus, the MSE is bounded as
\begin{equation}
  E(\hat\theta - \theta)^2  \le \prod_{i=1}^kr_i^{-2} + \epsilon\cdot 1.
\end{equation}
Set the value of $\epsilon$ in the protocol to $3^{-2k}$ and notice
the fact that $r_i\in\{2,3\}$, we have
\begin{equation}
  E(\hat\theta - \theta)^2 \le 2\prod_{i=1}^kr_i^{-2}.
\end{equation}

It is easy to prove by induction on $k$ that
\begin{equation}
  1 + r_1 + r_1r_2 + \cdots + r_1r_2\cdots r_{k-1} \le r_1r_2\cdots r_k.
\end{equation}
From this inequality, Eq.~\eqref{eq:total_number} and the fact that
both $n$ and $\log N$ is of order $k$, it follows that the RMSE of our
protocol is of order $O(\log N/N)$.

One can see that the protocol depends only on the ability to amplify
the parameter with linear cost before observing it. Thus it can be
applied to any problem where the amplification is possible. We will
see later a quite different use of this technique.

Besides the generality, the amplification protocol used here is
superior also in that it does not require the help of any
entanglement which is indispensable in the parallel strategy. It even
requires no joint measurement. This may make it easier to implement
experimentally than the other protocols we have mentioned.

As a conclusion of this section, we have discussed various strategies
that can beat the standard quantum limit in the case of estimating
parameter of a unitary operation. It is necessary to mention that, as
a generalization, it is proved recently that any unknown unitary can
be estimated efficiently of order $1/N$~\cite{Kahn06,RG03,CDPS04}.
Thus, all parameters of unitary operators possess a quite
non-classical property, the easiness of being estimated. However, we
will see that this is not always true for general quantum channels.

\section{Programmability and efficiency}
\label{sec:programmability}

In this section, we will begin to discuss the estimation of parameters
of a noisy quantum channel. We first give a formal definition of the
problem. As introduced in Section~\ref{sec:background}, a quantum
channel is a completely positive and trace-preserving superoperator. The
set of all quantum channels that maps density operators of
$\mathcal{H}_{in}$ to densities of $\mathcal{H}_{out}$ forms a
continuous manifold $\mathfrak{D}$. The unknown parameter belongs to
set $\Theta$, a continuous manifold of finite dimension. Finally, a
continuous injection $\mathcal{E}: \Theta \rightarrow \mathfrak{D}$
defines the family of parameterized quantum channels
$\{\mathcal{E}_\theta \mid \theta \in \Theta \}$.

\subsection{Programmable quantum channels}
\label{sec:programmable_channels}

The idea of programmable gates stems from the design of digital
circuits. It provides the convenience to change the functionality of a
gate by the control over some of its inputs. Theoretically, it is
possible to program all Boolean functions of $n$ bits into a single
gate. It is an interesting question to ask whether this is also
possible for quantum gates~\cite{NC97}. However, even the number of
all the quantum gates on single qubit is uncountably infinite.
Therefore, one cannot use classical controls to achieve this. But
does the use of quantum programs help here? Nielsen and
Chuang~\cite{NC97} gave a negative answer to this question.

\begin{figure}[!ht]
  \centering
  \includegraphics{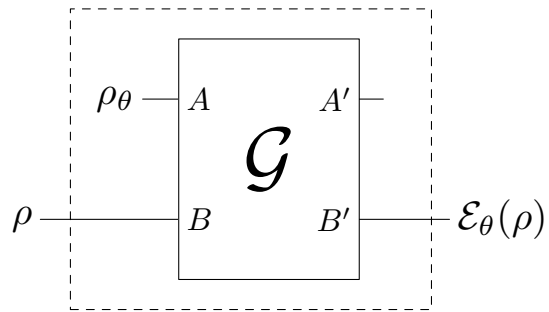}
  \caption{Illustration of the definition of programmable channels}
  \label{fig:programmable}
\end{figure}

To be precise, we define the notion of programmable gates as follows.
Let $\{\mathcal{E}_\theta\}$ be a family of quantum channels. It is
called programmable by $(\{\rho_\theta\},\mathcal{G})$ if there exist
a family of quantum states $\{\rho_\theta\}$ of a finite dimensional
space $\mathcal{H}$, and a quantum gate $\mathcal{G}$ that does not
depend on $\theta$ such that
\begin{equation}
  \label{eq:programmable_def}
  \mathcal{E}_\theta(\rho) = \tr_{A'} ( \mathcal{G} (\rho_\theta^A \otimes
  \rho^B) ),
\end{equation}
for all $\theta$ and $\rho$. In this definition, system $A$ stores the
quantum program and system $B$ receives the input data. The output
$B'$ is not necessarily equal to $B$. This definition is illustrated
in Fig.~\ref{fig:programmable}. One can always choose gate
$\mathcal{G}$ to be unitary without loss of generality. When
$\{\mathcal{E}_\theta\}$ is programmable we will also say that the
parameter $\theta$ is programmable.
\floatstyle{plain}
\restylefloat{figure}

What Nielsen and Chuang proved in~\cite{NC97} is that the family of
all unitary operations acting on $m$ qubits is not programmable. This
fact follows from the linearity of quantum mechanics. Fortunately,
they also pointed out the possibility to program the family of
unitaries in a probabilistic way. Vidal, Masanes and Cirac give a more
elegant construction to implement unitaries
probabilistically~\cite{VMC02}. What is more, the success probability
can be exponentially small when the size the quantum program grows. We
will present the protocol of Vidal \textit{et al.\/} in the following.
It is important to note that when we call a channel programmable, we
mean ``exact'' programmability as indicated by the above definition,
not the ``probabilistic'' compromise discussed here.

Generally, to prove that some family of channels is programmable, we
need to specify two things: the quantum program $\rho_\theta$ and the
quantum circuit $\mathcal{G}$. The former can be written out directly.
And we will describe the latter using the \textit{Quantum Computation
Language (QCL)} designed by \"{O}mer~\cite{Omer03}. The language has
a syntax derived from classical procedural languages like C or Pascal.
The main quantum features of it used in this paper are the quantum
data type \texttt{qureg}, which is an array of qubits, and the
statement \texttt{measure q[,var];}, which measures the register
\texttt{q} and assigns the result to the integer variable \texttt{var}
if specified. We will need one more statement \texttt{discard}, which
is not included in QCL, to represent the partial trace operation in
Eq.~\eqref{eq:programmable_def}. The comments interlaced in the codes
may help if one is not familiar with QCL.

Listed in Fig.~\ref{fig:probunitary} is the way of Vidal \textit{et
  al.\/} implementing
\begin{equation}
  U = \begin{bmatrix} e^{i\theta} &0\\0& e^{-i\theta} \end{bmatrix}
\end{equation}
with probability $1-2^{-k}$~\cite{VMC02}.

\floatstyle{boxed}
\restylefloat{figure}
\begin{figure*}[!htbp]
\caption{The probabilistic implementation of $U(\theta)$\label{fig:probunitary}}
\verb+  +The quantum program $\rho_\theta$, referred to in the code as
\verb+prog+, is chosen to be
\begin{equation*}
  \mathop{\bigotimes}\limits_{m=1}^k \bigl( e^{2^{m-1}\theta i}\ket{0} +
  e^{-2^{m-1}\theta i}\ket{1} \bigr),
\end{equation*}
\verb+  +and circuit $\mathcal{G}$ is specified in the following code:
\fontsize{10}{12}
\begin{verbatim}

  int probunitary(qureg prog, qureg in)
  {
    int r;
    int k = #prog;          // set k to the length of the program
    for i=0 to k-1 {
      CNot(prog[i],in[0]);  // apply the CNot gate, in[0] is the control bit
      measure prog[i],r;    // measure the ith qubit of prog, store the result in r
      if r==0 { return 1; } // success!
    }
    return 0;               // fail
  }
\end{verbatim}
\end{figure*}

\subsection{A no-go criterion and its applications}

We propose in the following a simple criterion that characterizes a
large class of inefficient parameters. The criterion is another
application of the quantum Cram\'{e}r-Rao inequality.

\begin{theorem}
\label{thm:programmable}
Any estimation of programmable parameter is of order $\Omega(1/\sqrt{N})$.
\end{theorem}

\begin{proof}
  Suppose $\{\mathcal{E}_\theta\}$ is programmable by
  $(\{\rho_\theta\},\mathcal{G})$. We need to show that any estimation
  of the parameter $\theta$ in $\mathcal{E}_\theta$ is of order
  $\Omega(1/\sqrt{N})$. The main point here is that any protocol
  $\mathcal{P}$ estimating $\theta$ of channel $\mathcal{E}_\theta$
  can be easily translated to a protocol $\mathcal{P}'$ that estimates
  the parameter of state $\rho_\theta$ with the same efficiency. The
  only difference in $\mathcal{P}$ and $\mathcal{P}'$ is that whenever
  $\mathcal{P}$ employs the channel $\mathcal{E}_\theta$, protocol
  $\mathcal{P}'$ takes a new copy of $\rho_\theta$ and applies gate
  $\mathcal{G}$. If the channel $\mathcal{E}_\theta$ is used $N$ times
  in protocol $\mathcal{P}$, $\mathcal{P}'$ will be a protocol
  estimating parameter $\theta$ using $N$ copies of $\rho_\theta$. As
  implied by the quantum Cram\'{e}r-Rao inequality, protocol
  $\mathcal{P}'$ is of order $\Omega(1/\sqrt{N})$, and so is protocol
  $\mathcal{P}$.
\end{proof}

We will now give some of the important applications of the criterion
stated in Theorem~\ref{thm:programmable}, beginning with simpler ones.

The first example considers the problem of estimating the noise level
of a qubit depolarizing channel parameterized by $\theta$. That is,
$\Theta = [0,1]$, and
\begin{equation}
  \label{eq:depolarizing}
  \mathcal{E}_\theta(\rho) = \theta {I \over 2} + (1-\theta) \rho.
\end{equation}
It is obvious that the larger the value of $\theta$, the noisier is
the channel. We note that the problem of optimally estimating
parameters of a depolarizing channel was studied in~\cite{SBB02}.

This family of channels is easily seen to be programmable. We list the
code in Fig.~\ref{fig:depolarizing}. An immediate consequence is that
parameter $\theta$ in Eq.~\eqref{eq:depolarizing} can never be
estimated superefficiently.

\floatstyle{boxed}
\restylefloat{figure}
\begin{figure*}[!htbp]
\caption{The programmable implementation of the depolarizing channel\label{fig:depolarizing}}
\verb+  +The quantum program, \verb+prog+, is chosen to be
\begin{equation*}
  (\sqrt{1-\theta} \ket{0} + \sqrt{\theta} \ket{1}) \otimes (\ket{00}
  + \ket{11})/\sqrt{2},
\end{equation*}
\verb+  +and the following code specifies the circuit:
\fontsize{10}{12}
\begin{verbatim}

  procedure depolarizing(qureg prog, qureg in)
  {
    int r;
    measure prog[0],r;      // measure the first qubit of prog, store the result in r
    if r==1 {
      Swap(prog[1],in[0]);  // apply the Swap gate
    }
    discard prog;           // trace out prog
  }
\end{verbatim}
\end{figure*}

To guarantee the channel defined in Eq.~\eqref{eq:depolarizing} to be
completely positive, $\theta$ can vary between $0$ and $4/3$.
Therefore, we can choose the parameter set to be a larger set
$\Theta=[0,4/3]$ and the parameter is still programmable. However, we
will not give the construction of it here because it is a special case
of the Pauli channels discussed below.

We have seen that the noise level of a depolarizing channel cannot be
estimated superefficiently. Similarly, any parameters that play the role of
probabilities, or functions related to probabilities, can be
programmed and are thus inefficient. Another example
falling in this type is the estimation of parameters of the Pauli
channel,
\begin{equation}
  \mathcal{E}_\theta (\rho) = \sum_{i=0}^3 p_i(\theta) \sigma_i \rho
  \sigma_i,
\end{equation}
where the $\sigma_i$'s are the Pauli matrices defined in
Eq.~\eqref{eq:pauli}. The programmable implementation of this family
is depicted in Fig.~\ref{fig:Pauli}.

\floatstyle{boxed}
\restylefloat{figure}
\begin{figure*}[!htbp]
\caption{The programmable implementation of the Pauli channel\label{fig:Pauli}}
\verb+  +The quantum program, \verb+prog+, is chosen to be
\begin{equation*}
  \sum_{i=0}^3 \sqrt{p_i} \ket{i},
\end{equation*}
\verb+  +and the following code specifies the circuit:
\fontsize{10}{12}
\begin{verbatim}

  procedure Pauli(qureg prog, qureg in)
  {
    int r;
    measure prog,r;         // measure prog and store the result in r
    if r==1 {
      X(in[0]);             // apply the X gate
    }
    if r==2 {
      Y(in[0]);             // apply the Y gate
    }
    if r==3 {
      Z(in[0]);             // apply the Z gate
    }
    discard prog;           // trace out prog
  }
\end{verbatim}
\end{figure*}

As mentioned, the family of depolarizing channels defined in
Eq.~\eqref{eq:depolarizing} is in fact the family of Pauli channels
with $p_0 = 1-3\theta/4$, $p_1=p_2=p_3=\theta/4$, so are the bit flip
and phase flip channels. The problem of parameter estimation of this
family is discussed in~\cite{FI03}. Using our criterion, we
immediately understand that no matter how hard we design the
estimator, the estimation protocol we can obtain is as efficient as
the most trivial one, tomography for example, up to some constant
factor.

There are some other simple examples that can be analyzed using
Theorem~\ref{thm:programmable}. For example, the channel
\begin{equation}
  \mathcal{E}_\theta (\rho) = \theta \oprod{0}{0} + (1-\theta) \rho,
\end{equation}
is programmable and thus $\theta$ is inefficient. Another similar
example is defined as
\begin{equation}
  \mathcal{E}_\theta (\rho) = \epsilon \oprod{\theta}{\theta} +
  (1-\epsilon) \rho,
\end{equation}
where $\epsilon$ is known and $\ket{\theta}$ is some pure state
depending on the parameter.

The only family of qubit channels of common interests whose estimation
efficiency cannot be characterized using
Theorem~\ref{thm:programmable} is the amplitude damping channels. The
following Kraus' operators give the parametrization:
\begin{equation}
  E_0 = \begin{bmatrix}1&0\\0&\sqrt{1-\theta}\end{bmatrix},\quad
  E_1 = \begin{bmatrix}0&\sqrt{\theta}\\0&0\end{bmatrix}.
\end{equation}
It is proved in~\cite{HZB02} that the family of amplitude damping
channels is not programmable. Therefore,
Theorem~\ref{thm:programmable} does not apply any more. Interestingly,
however, the so called generalized amplitude damping channels having
Kraus' operators as
\begin{equation}
\begin{split}
  E_0 & = \sqrt{p} \begin{bmatrix}1&0\\0&\sqrt{1-\theta}\end{bmatrix},\quad
  E_1 = \sqrt{p} \begin{bmatrix}0&\sqrt{\theta}\\0&0\end{bmatrix},\\
  E_2 & = \sqrt{1-p} \begin{bmatrix}\sqrt{1-\theta}&0\\0&1\end{bmatrix},\quad
  E_3 = \sqrt{1-p} \begin{bmatrix}0&0\\\sqrt{\theta}&0\end{bmatrix},
\end{split}
\end{equation}
are programmable for any fixed $p\in (0,1)$. We will prove it in
Section \ref{sec:depolarizing_ruins_efficiency}.

\subsection{Classical channels are all programmable}

We will show in the following that classical information channels are
quantum programmable. Interestingly, this gives a simple proof for the
fact that parameters of any family of classical channels cannot be
estimated with a convergence rate better than $1/\sqrt{N}$.

A discrete memoryless channel is characterized by a set of transition
probabilities $p(y|x)$ which define an $m$ by $n$ stochastic matrix
$Q=(q_{xy})$. Just like any classical computation can be thought of as
a special quantum computation, we can also regard the DMC $Q$ as a
quantum channel $\mathcal{Q}$. Channel $\mathcal{Q}$ will assume
restricted inputs, namely, diagonal density matrices, and will
guarantee the output to be diagonal densities too.

The main idea of the programmable construction of $\mathcal{Q}$ is to
simulate the transition probabilities after measuring the input in the
computational basis. The construction is given in Fig.~\ref{fig:dmc}.

\floatstyle{boxed}
\restylefloat{figure}
\begin{figure*}[!htbp]
\caption{The programmable implementation of classical DMC.\label{fig:dmc}}
\verb+  +Assume without loss of generality that $n=2^k$. The quantum
program, \verb+prog+, is chosen to be
\begin{equation*}
  \left[ \mathop{\bigotimes}\limits_{x=0}^{m-1} \sum_{y=0}^{n-1}
    \sqrt{q_{xy}} \ket{y} \right] \otimes \ket{0},
\end{equation*}
\verb+  +whose last part consists of $k$ qubits and the following code
specifies the circuit:
\fontsize{10}{12}
\begin{verbatim}

  procedure dmc(qureg prog, qureg in)
  {
    int r;
    measure in,r;           // measure the input and store the result in r
    measure prog[r*k::k];   // measure the subregister of length k started at r*k
    Swap(prog[r*k::k],prog[m*k::k]);
                            // Swap two subregisters of prog
    discard prog[0::m*k];   // prog[m*k::k], the last part of prog, remains
                            // and will be the output
    discard in;
  }

\end{verbatim}
\end{figure*}

Thus, it follows from Theorem~\ref{thm:programmable} that any
parameter of a DMC cannot be estimated superefficiently. This is in
contrast to the quantum case. Table~\ref{tab:efficient} illustrates
the interesting circumstances of whether a parameter is superefficient or
not.

\begin{table}[!ht]
  \renewcommand\arraystretch{1.5}
  \centering\large
  \begin{tabular}{|r||c|c|}
  \hline
   & Source & Channel\\
  \hline
  Classical & NO & NO\\
  \hline
  Quantum & NO & Possibly YES\\
  \hline
  \end{tabular}
  \caption{Superefficient parameters}
  \label{tab:efficient}
\end{table}

\subsection{Depolarizing noise ruins the efficiency}
\label{sec:depolarizing_ruins_efficiency}

Our last application of the ``no-go'' criterion considers the family
of quantum channels that has the following form:
\begin{equation}
  \label{eq:dnoise_form}
  \mathcal{E}_\theta (\rho) = \epsilon \rho_0 + (1-\epsilon)
  \mathcal{U}_\theta(\rho),
\end{equation}
where $\epsilon\in(0,1]$, $\rho_0$ is independent of $\theta$ and
$\mathcal{U}_\theta$ is another family of channels. Let
$\mathcal{U}_\theta$ be a family of unitaries first. For example,
\begin{equation}
  \mathcal{U}_\theta(\rho) = U_\theta \rho U_\theta^{\dagger},\quad
  U_\theta = \begin{bmatrix}e^{i\theta} &0\\0&e^{-i\theta}\end{bmatrix}.
\end{equation}

When $\rho_0$ is $I/d$, the channel $\mathcal{E}_\theta$ is an
imperfect implementation of $\mathcal{U}_\theta$ disturbed by the
depolarizing noise. If $\epsilon$ is close to $0$, then
$\mathcal{E}_\theta$ is intuitively also close to
$\mathcal{U}_\theta$. As we have seen that $\theta$ in
$\mathcal{U}_\theta$ can estimated of order $1/N$, we may expect that
$\theta$ of $\mathcal{E}_\theta$ can also be estimated
superefficiently. However, we will see that no matter how small
$\epsilon$ is, this cannot happen. That is, the depolarizing noise has
totally ruined the efficiency of estimations!

The proof is simple. As we have shown how to program unitary operation
with arbitrarily high probability in
Section~\ref{sec:programmable_channels}, we can construct the
programmable realization of $\mathcal{E}_\theta$ easily: implement
$\mathcal{U}_\theta$ probabilistically and output $\rho_0$ on failure.
See Fig.~\ref{fig:dnoise} for details.

\floatstyle{boxed}
\restylefloat{figure}
\begin{figure*}[!htbp]
\caption{The programmable implementation of $\mathcal{E}_\theta$\label{fig:dnoise}}
\verb+  +Choose $k$ such that $\epsilon \ge 2^{-k}$. The quantum
program, \verb+prog+, is chosen to be
\begin{equation*}
  \mathop{\bigotimes}\limits_{m=1}^k \bigl( e^{2^{m-1}\theta i}\ket{0} +
  e^{-2^{m-1}\theta i}\ket{1} \bigr) \otimes \rho_0 \otimes
  \bigl(\oprod{1}{1} + (\epsilon-2^{-k})Z \bigr),
\end{equation*}
\verb+  +and the following code specifies the circuit:
\fontsize{10}{12}
\begin{verbatim}

  procedure dnoise(qureg prog, qureg in)
  {
    int r;
    measure prog[k+1],r;    // measure the last qubit of prog, store the result in r
    if probunitary(prog[0::k], in)==0 or r==0 {
      Swap(prog[k], in[0]); // Swap (with probability \epsilon)
    }
    discard prog;
  }
\end{verbatim}
\end{figure*}

The above negative result casts shadow on all of the fast estimation
protocols proposed so far. As a small amount of depolarizing noise is
unavoidable when conducting experiments of estimation, it seems that
we will not be able to beat the standard quantum limit in practice.
However, as long as the noise is indeed small and the size $N$ is not
too large to amplify the noise to a noticeable magnitude, it will
still be possible to exploit the quantum advantage in the fast
estimation protocols~\cite{BB05}. That is, there is a trade-off
between the noise level and the sample size to preserve the speedup.
We analyze the modified bitwise estimation protocol as a
demonstration.

Consider the family of channels
\begin{equation}
  \mathcal{E}_\theta (\rho) = \epsilon \frac{I}{2} + (1-\epsilon)
  U_\theta\rho U_\theta^\dagger, 
\end{equation}
with $U_\theta$ defined in Eq.~\eqref{eq:single_phase_2}:
\begin{equation}
  \begin{bmatrix}1 & 0\\0 & e^{2\pi i\theta}\end{bmatrix}.\nonumber
\end{equation}
We will see that the scaling of $\log N/N$ is preserved if
\begin{equation}
  N \epsilon \le 1,
\end{equation}
which represents the trade-off rigorously. To see this, notice by
induction on $m$ that
\begin{equation}
  \mathcal{E}_\theta^m = \bigl(1 - (1-\epsilon)^m\bigr) \frac{I}{2} +
  (1-\epsilon)^m U_{m\theta}\rho U_{m\theta}^\dagger.
\end{equation}
It implies that the corresponding probability $P_+'$ and $P_-'$ are
related to their noiseless counterpart as
\begin{equation}
\begin{split}
  P_+' & = \frac{1-(1-\epsilon)^m}{2} + (1-\epsilon)^m P_+\\
  P_-' & = \frac{1-(1-\epsilon)^m}{2} + (1-\epsilon)^m P_-.
\end{split}
\end{equation}
As $m$ is at most $\prod r_i = O(N/\log N)$ in the protocol, we have
\begin{equation}
  (1-\epsilon)^m \ge (1-\frac{1}{N})^m \ge 1 - \frac{m}{N} \ge 1 -
  O(1/\log N).
\end{equation}
That is, $(1-\epsilon)^m$ is close to $1$ for large $N$, and thus
$P_\pm'$ are also close to $P_\pm$. The efficiency of order $O(\log
N/N)$ remains as claimed.

As it is proved in~\cite{VMC02} that any unitary operation can be
programmed with arbitrarily high probability, one can easily see from
Eq.~\eqref{eq:traced_u} that any quantum channel can also be programmed
probabilistically. Thus, we can easily generalize the result to
quantum channels of form Eq.~\eqref{eq:dnoise_form} where
$\mathcal{U}_\theta$ are on longer restricted to unitaries.

In the rest of this section, we give a more direct characterization of
channels suffering from depolarizing noise and use it to discuss the
estimation problem of the generalized amplitude damping channels.

It is well known that there is a correspondence between a quantum
channel $\mathcal{E}$ and its state representative (a.k.a. Choi
matrix) $\mathcal{I}\otimes\mathcal{E} (\oprod{\Psi}{\Psi})$, where
$\ket{\Psi}$ is the maximally entangled state. Superoperator
$\mathcal{E}$ is complete positive if and only if its Choi matrix is
positive semidefinite~\cite{NC00}. Using this correspondence, it not
difficult to show the following Lemma and we omit its proof.

\begin{lemma}
  Let $\mathcal{E}$ be a quantum channel with a Choi matrix
  $\mathcal{C}$. There exists some constant $\epsilon\in(0,1)$ and quantum
  channel $\mathcal{U}$ such that
\begin{equation}
  \mathcal{E} (\rho) = \epsilon \rho_0  + (1-\epsilon)
  \mathcal{U}(\rho),
\end{equation}
if and only if the null space of $\mathcal{C}$ is contained in that of
$\oprod{+}{+} \otimes \rho_0$, where $\ket{+}$ is state
$\sum_{i=0}^{d-1}\ket{i}/\sqrt{d}$.
\end{lemma}

An immediate corollary of the Lemma is that when the Choi matrix of a
channel $\mathcal{E}$ is positive definite, $\mathcal{E}$ suffers from
depolarizing noise. We use this fact to show the programmability of
generalized amplitude damping channel for fixed $p\in (0,1)$ and
parameter set $\Theta = [a,1]$, $a>0$. The Choi matrix of the
generalized amplitude damping channel is
\begin{equation}\frac{1}{2}
  \begin{bmatrix}
    1-\theta+p\theta&0&0&\sqrt{1-\theta}\\
    0&\theta-p\theta&0&0\\
    0&0&p\theta&0\\
    \sqrt{1-\theta}&0&0&1-p\theta
  \end{bmatrix}
\end{equation}
whose eigenvalues are $p\theta/2,(\theta-p\theta)/2$ and
\begin{equation}
  \frac{2-\theta \pm \sqrt{(2-\theta)^2-4p(1-p)\theta^2}}{4}.
\end{equation}
It is easy to verify that all these eigenvalues are strictly positive.
The programmability therefore follows from the above Lemma and the
fact that channels suffers from depolarizing noise are programmable.

\section{Fast estimation of parameters in noisy channels}
\label{sec:projector_class}

Up to now, we have seen that parameters of unitary operation can be
estimated superefficiently. We have also shown many examples of
quantum channels in which superefficient parameter estimation is
impossible by the ``no-go'' criterion we provide. We will now answer
the question of whether fast parameter estimation can occur in
non-unitary channels, or it is only a unique phenomenon for unitaries.
Of course, there are trivial positive examples:
$\mathcal{E}_\theta(\rho) = \frac{I}{2} \otimes
\mathcal{U}_\theta(\rho)$ where $\frac{I}{2}$ is maximally mixed state
and $\mathcal{U}_\theta$ is unitary. What we will discuss in the
following is not of this kind.

We first give the family of channels of interest and the protocol to
estimate the parameter. Define an intermediate matrix $E$ as the qubit
rotation
\begin{equation}
  \label{eq:E}
  \renewcommand\arraystretch{1.5}
  \begin{split}
  E & = \cos\theta I + i \sin\theta Y\\
    & = \begin{bmatrix} \phantom{-}\cos\theta & \sin\theta\\
     -\sin\theta & \cos\theta \end{bmatrix},
  \end{split}
\end{equation}
where $\theta\in[0,\pi/2]$ and define $E_0$ and $E_1$ in terms of $E$ as
\begin{equation}
  E_0 = \sqrt{\eta_\theta} E,\quad E_1 = \sqrt{1-\eta_\theta} Z E.
\end{equation}
Here, $\eta_\theta$ is any real function of $\theta$ that ranges in
$[0,1]$ and $Z$ is one of the Pauli matrices defined in
Eq.~\eqref{eq:pauli}.

Now, we can write out the family of channels explicitly using the
operator-sum representation:
\begin{equation}
  \label{eq:projector_class_channel}
  \mathcal{E}_\theta (\rho) = E_0\rho E_0^{\dagger} + E_1\rho E_1^{\dagger},
\end{equation}
with $\theta\in\Theta=[0,\pi/2]$.

When $\eta_\theta = 1/2$, a constant function, this family of quantum
channels is a collection of single-qubit observables, which is indeed
far from unitary evolutions as claimed. Specifically, it easy to check
that, in this case,
\begin{equation}
\begin{split}
  \oprod{0}{\theta_0} & = \phantom{-}1/\sqrt{2} E_0 + 1/\sqrt{2} E_1\\
  \oprod{1}{\theta_1} & = -1/\sqrt{2} E_0 + 1/\sqrt{2} E_1
\end{split}
\end{equation}
where $\ket{\theta_0} = \cos\theta\ket{0} + \sin\theta\ket{1}$
and $\ket{\theta_1} = \sin\theta\ket{0} - \cos\theta\ket{1}$.
As the matrix
\begin{equation}
  \frac{1}{\sqrt{2}}\begin{bmatrix}1&1\\-1&1\end{bmatrix}
\end{equation}
is unitary, it follows from the unitary freedom of operator-sum
representation that, when $\eta_\theta = 1/2$, we can write
\begin{equation}
  \mathcal{E}_\theta (\rho) = \oprod{0}{\theta_0} \rho
  \oprod{\theta_0}{0} + \oprod{1}{\theta_1} \rho \oprod{\theta_1}{1},
\end{equation}
an operator-sum representation of projective measurement along the
basis $\{\ket{\theta_0},\ket{\theta_1}\}$. We will thus call the
family defined in Eq.~\eqref{eq:projector_class_channel} the
\textit{projector-class channels\/}. Denote by $\mathcal{M}_\theta$ the
special case of the family of $\eta_\theta = 1/2$ and denote by
$\mathcal{M} = \mathcal{M}_0$ the measurement along computational
basis. Our result can be summarized in the following theorem.

\begin{theorem}
  Parameter $\theta$ of the projector-class channels is of order
  $O(\log N/N)$.
\end{theorem}

To prove the theorem, We will specify the protocol by employing the
amplification technique introduced in Section~\ref{sec:amplify}. It is
only necessary to construct a distribution related to the amplified
parameter $n\theta$ by $n$ uses of the channel. We will achieve it in
two steps.

First, prepare an $n$-qubit entangled state
\begin{equation}
  \ket{\Psi_n} = \sum_{i\in E_n} (-1)^{w(i)/2}\ket{i}/\sqrt{2^{n-1}},
\end{equation}
where $E_n$ is the set of all $n$-bit strings of $0$ and $1$ with even
parity and $w(i)$ is the Hamming weight of $i$. This state is first
used in~\cite{JFDY06} to identify observables by the authors.

Next, apply the channel $\mathcal{E}_\theta$ on each qubit of the
state and measure in the computational basis.

We now calculate the probability of measurement outcomes having an
even parity $\Pr(even)$. The following identities can simplify the
analysis:
\begin{equation}
  \mathcal{M} \circ \mathcal{E}_\theta = \mathcal{M}_\theta =
  \mathcal{M} \circ \mathcal{U}_\theta,
\end{equation}
where $\mathcal{U}_\theta$ corresponds to the unitary operation
$\oprod{0}{\theta_0}+\oprod{1}{\theta_1}$. The second equality is
obvious and we prove the first one only. $\mathcal{M} \circ
\mathcal{E}_\theta$ has a representation consisting of the follow four
operators:
\begin{equation}
\begin{split}
  \oprod{0}{0} E_0 & = \sqrt{\eta_\theta} \oprod{0}{\theta_0}\\
  \oprod{0}{0} E_1 & = \sqrt{1-\eta_\theta} \oprod{0}{\theta_0}\\
  \oprod{1}{1} E_0 & = -\sqrt{\eta_\theta} \oprod{1}{\theta_1}\\
  \oprod{1}{1} E_1 & = \sqrt{1-\eta_\theta} \oprod{1}{\theta_1}.
\end{split}
\end{equation}
The first two terms can be merged to one of the operators of
$\mathcal{M}_\theta$, so is the second two. The equality is thus
proved.

An implication of this identity is that $\Pr(even)$ remains the
same if we substitute $\mathcal{U}_\theta$ for $\mathcal{E}_\theta$ in
the second step of the above protocol. That is,
\begin{equation}\label{eq:prob_even}
\begin{split}
  \Pr(even) & = \sum_{j\in E_n} \bra{j}
  \mathcal{U}_\theta^{\otimes n}(\oprod{\Psi_n}{\Psi_n}) \ket{j}\\
            & = \sum_{j\in E_n} \bigl|\bra{j}U^{\otimes n}\ket{\Psi_n}\bigr|^2,
\end{split}
\end{equation}
where $U = \oprod{0}{\theta_0}+\oprod{1}{\theta_1}$.

The element of the $j$th row and $k$ column of $U^{\otimes n}$ is
\begin{equation}
  (-1)^{w(j\cdot k)} (\cos\theta)^{n-d(j,k)} (\sin\theta)^{d(j,k)},
\end{equation}
where $d(j,k)$ is the Hamming distance function and $j\cdot k$ is the
bitwise AND of $j,k$. Consequently, $\bra{j}U^{\otimes n}\ket{\Psi_n}$
equals to
\begin{equation}
  \frac{1}{\sqrt{2^{n-1}}}
  \sum_{k\in E_n} (-1)^{w(k)/2} (-1)^{w(j\cdot k)}
  (\cos\theta)^{n-d(j,k)}(\sin\theta)^{d(j,k)}.
\end{equation}
Notice the fact that $2w(j\cdot k) + d(j,k) = w(j) + w(k)$, the above
quantity can be further simplified as
\begin{equation}
  \frac{(-1)^{w(j)/2}}{\sqrt{2^{n-1}}} \sum_{k\in E_n}
  (-1)^{-d(j,k)/2} (\cos\theta)^{n-d(j,k)}(\sin\theta)^{d(j,k)}.
\end{equation}
We can write the summation without the multiplicative constant $(-1)^{w(j)/2}/\sqrt{2^{n-1}}$ as
\begin{equation}
  \sum_{k\in E_n} (\cos\theta)^{n-d(j,k)}(-i\sin\theta)^{d(j,k)}.
\end{equation}
For every even $l$, there are exactly ${n\choose l}$ different k's such that $d(j,k) = l$. Therefore, the summation is
\begin{equation}
\begin{split}
    & \sum_{l \text{ is even}} {n\choose l}\cos^{n-l}\theta(-i\sin\theta)^{l}\\
  = & \re \sum_{l=0}^n{n\choose l}\cos^{n-l}\theta(-i\sin\theta)^{l}\\
  = & \re e^{-in\theta} = \cos (n\theta).
\end{split}
\end{equation}

Taking the constant into account, we have
\begin{equation}
  \bra{j}U^{\otimes n}\ket{\Psi_n} = 
  \frac{(-1)^{w(j)/2}}{\sqrt{2^{n-1}}} \cos (n\theta),
\end{equation}
which enables us to finish the calculation of $\Pr (even)$ in Eq.~\eqref{eq:prob_even} as
\begin{equation}
  \Pr (even) = \sum_{j\in E_n} \bigl|\bra{j}U^{\otimes
    n}\ket{\Psi_n}\bigr|^2 = \cos^2(n\theta).
\end{equation}

The parameter is thus amplified as promised with the help of the GHZ
entangled state. We need only to follow the idea of the modified
bitwise estimation protocol discussed in Section~\ref{sec:amplify}.
Clearly, the estimator is of order $O(\log N/N)$. We do not know
whether it is possible or not to achieve the order of $1/N$ in this
problem as in the case of estimating unitary operations.

\section{Conclusions}

In this paper, we have discussed the estimation theory of parameters
of quantum channels with emphasis on evaluating the efficiency of
estimation protocols. It is clear now that there are two fundamentally
different types of parameters of quantum channels, one of which can be
estimated superefficiently and the other cannot. The fact that all
programmable parameters are inefficient provides us with an
easy-to-use yet powerful way of determining whether superefficient
estimation is possible. Based on this fact, we have shown many
examples of inefficient parameters. What is more, it also follows that
all parameters of classical information channels are inefficient and
that depolarizing noise will undermine the efficiency universally. We
have also constructed the so-called projector-class channels and
provide an superefficient protocol to estimate the parameter of this
family. Thus superefficient estimation is not a unique phenomenon of
unitary operations. What remains valuable for further investigation in
future work is to evaluate the power of the ``no-go'' criterion in
terms of programmability, and to characterize, in a more direct way,
parameters of a quantum channel that can be superefficiently estimated.

\bibliographystyle{IEEEtran}
\bibliography{IEEEabrv,pares}

\end{document}